\documentclass[prd,aps,twoside,eqsecnum]{revtex4-1}

\usepackage{hyperref,amsfonts,amsmath,amssymb,graphicx,bm,subfigure}

\begin{document}

\author{Maxim Dvornikov}
\email{maxdvo@izmiran.ru}

\title{Neutrino spin oscillations in external fields in curved spacetime}

\affiliation{Pushkov Institute of Terrestrial Magnetism, Ionosphere
and Radiowave Propagation (IZMIRAN),
108840 Moscow, Troitsk, Russia}

%
%
%
%
%
%

\begin{abstract}
We study spin oscillations of massive Dirac neutrinos in background
matter, electromagnetic and gravitational fields. First, using the
Dirac equation for a neutrino interacting with the external fields
in curved spacetime, we rederive the quasiclassical equation for
the neutrino spin evolution, which was proposed previously basing
on principles of the general covariance. Then, we apply this result
for the description of neutrino spin oscillations in nonmoving and
unpolarized matter under the influence of a constant transverse magnetic
field and a gravitational wave. We derive the effective Schr\"odinger
equation for neutrino oscillations in these external fields and solve
it numerically. Choosing realistic parameters of external fields, we show that the parametric resonance can take place in spin oscillations of low energy neutrinos. Some astrophysical applications are briefly
discussed.
\end{abstract}


\maketitle

\section{Introduction}

Astrophysical neutrinos play an important role for the evolution of stars,
supernovae, and the early Universe~\cite{Raf96}. Neutrinos have
a remarkable property of transitions from one type to another, called
neutrino oscillations~\cite{Bil18}. Some external backgrounds can affect neutrino oscillations. For example, if the density of background matter changes adiabatically along the neutrino trajectory, the probability of flavor
oscillations $\nu_{e}\to\nu_{\mu}$ of neutrinos, interacting with this matter, can be amplified. This phenomenon is known as the Mikheyev-Smirnov-Wolfenstein
(MSW) effect. It is believed to be the most plausible solution to the solar neutrino
problem~\cite{MalSmi16}.

There are various types of neutrino oscillations (see, e.g., Ref.~\cite{Zub04}
for a review). Among them, we can discuss the possibility of transitions
between left and right polarized neutrinos of one type, $\nu_{-}\to\nu_{+}$,
neglecting the neutrino mixing. This process, named neutrino spin
oscillations, results in the effective reduction of an initial flux
of left polarized particles since the standard model interactions of right polarized neutrinos with other particles are strongly suppressed. One can describe neutrino spin oscillations
as the precession of the neutrino spin in an external field.

Besides the MSW effect, mentioned above, other external fields can influence
neutrino oscillations. For example, the gravitational interaction,
in spite of its weakness, was found in Ref.~\cite{AhlBur96} to contribute
to oscillations of neutrinos. In the present work, we are interested
in the influence of gravitational fields on neutrino spin oscillations.

The problem of the evolution of a spinning particle in general relativity
was tackled for the first time in Ref.~\cite{Pap51}. The resulting
evolution equations turn out to be nonlinear, i.e. the particle trajectory
is affected by the spin evolution and vice versa. Fortunately, in the
case of a pointlike elementary particle, such as a neutrino, the
particle motion is not influenced by the particle spin. It means that, 
with a high level of accuracy, 
neutrinos follow geodesic lines in an external gravitational field.

The equation for the quasiclassical description of the particle spin
evolution in a gravitational field was proposed in Ref.~\cite{PomKri98}.
Then, in Ref.~\cite{Dvo06}, this approach was adapted for the studies
of neutrino oscillations in a gravitational field. Neutrino oscillations
in matter under the influence of an electromagnetic field in curved
spacetime were studied in Ref.~\cite{Dvo13} by proposing the generalization
of the quasiclassical covariant equation for the neutrino spin evolution
in these external fields. The formalism, developed in Refs.~\cite{Dvo06,Dvo13},
was applied in Refs.~\cite{AlaNod15,Cha15} to study neutrino
spin oscillations in various gravitational backgrounds. The method
for the description of the particle spin evolution in a gravitational
field, based on the analysis of the Dirac equation in curved spacetime,
was recently developed in Ref.~\cite{ObuSilTer09}.

In the present work, we continue the study of neutrino spin
oscillations in external fields in curved spacetime. The quasiclassical
equation, accounting for the contribution of external fields to the
neutrino spin evolution, was derived in Ref.~\cite{Dvo13} based
on principles of the general covariance. However this approach has to
be substantiated by a more rigorous derivation. It deals mainly with
the contribution of background matter to the neutrino spin evolution
in curved spacetime. It is the main motivation of the present work.

Then, in Refs.~\cite{Dvo06,Dvo13}, we considered neutrino spin oscillations
in static gravitational backgrounds, like Schwarzschild and Kerr metrics.
It is interesting to analyze the dynamics of neutrino oscillations
in a time dependent metric induced, e.g., by a gravitational wave (GW). Typically astrophysical neutrinos
interact with background matter and an external magnetic field besides a gravitational field. The contributions to neutrino spin oscillations of all these external fields---namely, a background matter, a magnetic field, and a gravitational field (GW)---are accounted for in our study. Note that the evolution
of a fermion spin in GW was recently studied in Refs.~\cite{Qua16,ObuSilTer17}. The interest in the studies of neutrino oscillations in periodically varying external fields is inspired,
e.g., by the suggestion in Ref.~\cite{Fom18} to explore
the neutrino evolution in intense laser pulses.

The recent observation of GW by the LIGO and Virgo detectors~\cite{Abb16}
is the strong evidence of the validity of the general relativity.
The most powerful sources of GW are systems of binary black holes
(BHs)~\cite{Abb18} and neutron stars (NSs)~\cite{Abb17NS} during their coalescence. The merging of two compact objects can
be a source of both GW and other elementary particles including
neutrinos. Moreover, dense background matter and strong magnetic fields
are ejected in outer space in the coalescence of BHs or NSs. These
factors can affect the propagation and oscillations of emitted neutrinos.
There are numerous attempts to carry out multimessenger observations
of both GW and astrophysical neutrinos by the modern neutrino telescopes
and cosmic ray experiments~\cite{Alb19}. Recently the discovery of GWs was followed by the observation of the event horizon silhouette of the supermassive BH, reported in Ref.~\cite{Aki19}, which is another direct confirmation of the validity of the general relativity.


This paper is organized as follows. We start in Sec.~\ref{sec:FORMAL}
with the derivation of the quasiclassical equation for the neutrino
spin evolution in external fields in curved spacetime based on the
Dirac equation for a massive neutrino accounting for these external
fields. Then, in Sec.~\ref{sec:GW}, we apply our results for the
description of the neutrino spin evolution in background matter, a
transverse magnetic field, and GW. We demonstrate that a parametric
resonance can happen in oscillations of low energy neutrinos in realistic astrophysical backgrounds.
Finally, in Sec.~\ref{sec:DISC}, we discuss our results.

\section{Formalism for the description of the neutrino spin evolution\label{sec:FORMAL}}

In this section, we derive the covariant quasiclassical equation for
the neutrino spin evolution in external fields in curved spacetime.
For this purpose we analyze the corresponding Dirac equation. Our
result is compared with previous findings.

We consider one neutrino eigenstate, which is supposed to be a Dirac
particle, and neglect the mixing between different neutrino types.
The wave equation for a massive Dirac neutrino with the anomalous
magnetic moment, interacting with background matter and the electromagnetic
field $F_{\mu\nu}$ in curved spacetime, reads
\begin{equation}\label{eq:Direq}
  \left[
    \mathrm{i}\gamma^{\mu}\nabla_{\mu}-\frac{\mu}{2}F_{\mu\nu}\sigma^{\mu\nu}- 
    \frac{V^{\mu}}{2}\gamma_{\mu}
    \left(
      1-\gamma^{5}
    \right) -
    m
  \right]
  \psi = 0,
\end{equation}
where $\gamma^{\mu}=\gamma^{\mu}(x)$, $\sigma_{\mu\nu}=\tfrac{\mathrm{i}}{2}\left[\gamma_{\mu},\gamma_{\nu}\right]_{-}$,
and $\gamma^{5}=-\tfrac{\mathrm{i}}{4!}E^{\mu\nu\alpha\beta}\gamma_{\mu}\gamma_{\nu}\gamma_{\alpha}\gamma_{\beta}$
are the coordinate dependent Dirac matrices, $E^{\mu\nu\alpha\beta}=\varepsilon^{\mu\nu\alpha\beta}/\sqrt{-g}$
is the covariant antisymmetric tensor in curved spacetime, $g=\text{det}(g_{\mu\nu})$,
$g_{\mu\nu}$ is the metric tensor, $\nabla_{\mu}=\partial_{\mu}+\Gamma_{\mu}$
is the covariant derivative, $\Gamma_{\mu}$ is the spin connection,
$\mu$ is the magnetic moment of a neutrino, and $m$ is the neutrino
mass.

The effective potential $V^{\mu} = (V^0,\mathbf{V})$ of the neutrino interaction with
arbitrarily polarized and moving matter has the form,
\begin{equation}\label{eq:Vmudef}
  V^{\mu}=\sqrt{2}G_{\mathrm{F}}\sum_{f}
  \left(
    q_{f}^{(1)}j_{f}^{\mu}+q_{f}^{(2)}\lambda_{f}^{\mu}
  \right),
\end{equation}
where $j_{f}^{\mu}$ and $\lambda_{f}^{\mu}$ are the hydrodynamic
currents and the polarizations of background fermions of the type
$f$, $G_{\mathrm{F}}=1.17\times10^{-5}\,\text{GeV}^{-2}$ is the
Fermi constant, and $q_{f}^{(1,2)}$ are the constants that are given
in the explicit form in Ref.~\cite{DvoStu02}. The contribution of
the electroweak interaction of a fermion with background matter to
the Dirac equation in curved spacetime, as in Eq.~(\ref{eq:Direq}),
was previously considered in Refs.~\cite{Dvo14,Dvo15a,Dvo15b}.

To find the spin connection in Eq.~(\ref{eq:Direq}) we choose a
locally Minkowskian frame $x^{\mu}\to\bar{x}^{a}$: $\eta_{ab}=e_{a}^{\ \mu}e_{b}^{\ \nu}g_{\mu\nu}$,
where $e_{a}^{\ \mu}=\partial x^{\mu}/\partial\bar{x}^{a}$ are the
vierbein vectors and $\eta_{ab}=\text{diag}(+1,-1,-1,-1)$ is the
metric tensor in Minkowski space. In such a frame, $\Gamma_{\mu}$
is defined as $\Gamma_{\mu}=-\tfrac{\mathrm{i}}{4}\sigma^{ab}\omega_{ab\mu}$,
where $\omega_{ab\mu}=e_{a}^{\ \nu}e_{b\nu;\mu}$ are the components
of the connection one-form, $\bar{\sigma}_{ab}=\tfrac{\mathrm{i}}{2}\left[\bar{\gamma}_{a},\bar{\gamma}_{b}\right]_{-}$,
$\bar{\gamma}^{a}=e_{\ \mu}^{a}\gamma^{\mu}$ are the constant Dirac
matrices given in the chosen frame, and the semicolon stays for a
covariant derivative.

The Dirac operator, expressed in the locally Minkowskian coordinates $\bar{x}^a$, takes the form, $\mathrm{i}\gamma^{\mu}\nabla_{\mu}=\mathrm{i}\bar{\gamma}^{a}\partial_{a}+\frac{\mathrm{i}}{4}\bar{\gamma}^{a}\bar{\gamma}^{b}\bar{\gamma}^{c}\gamma_{cba},$
where $\gamma_{cba}=e_{c\mu;\nu}e_{b}^{\ \mu}e_{a}^{\ \nu}=-\gamma_{bca}$
are the Ricci rotation coefficients. Then, using the identity,
\begin{equation}\label{eq:3Dm}
  \bar{\gamma}^{a}\bar{\gamma}^{b}\bar{\gamma}^{c}=
  \eta^{ab}\bar{\gamma}^{c}+\eta^{bc}\bar{\gamma}^{a}-\eta^{ac}\bar{\gamma}^{b}-
  \mathrm{i}\varepsilon^{dabc}\bar{\gamma}_{d}\bar{\gamma}^{5},
\end{equation}
where $\varepsilon^{abcd}$ is the absolute antisymmetric tensor in
Minkowski space, having $\varepsilon^{0123}=+1$, and $\bar{\gamma}^{5}=\mathrm{i}\bar{\gamma}^{0}\bar{\gamma}^{1}\bar{\gamma}^{2}\bar{\gamma}^{3}$,
we rewrite the Dirac Eq.~(\ref{eq:Direq}) in the form,
\begin{equation}\label{eq:Direqvf}
  \left[
    \bar{\gamma}^{a}
    \left(
      \mathrm{i}\partial_{a}+\frac{\mathrm{i}}{2}\gamma_{abc}\eta^{bc}
    \right) +
    \bar{\gamma}^{a}\bar{\gamma}^{5}\frac{1}{4}\varepsilon_{abcd}\gamma^{cbd}-
    \frac{\mu}{2}f_{ab}\bar{\sigma}^{ab}-\frac{v^{a}}{2}\bar{\gamma}_{a}
    \left(
      1-\bar{\gamma}^{5}
    \right) -
    m
  \right]
  \psi=0.
\end{equation}
Here $f_{ab}=e_{a}^{\ \mu}e_{b}^{\ \nu}F_{\mu\nu}$ and $v^{a}=e_{\ \mu}^{a}V^{\mu}$
are the corresponding objects expressed in the locally Minkowskian frame.

The axial-vector contribution $\sim\bar{\gamma}^{a}\bar{\gamma}^{5}$ of
a gravitational field to Eq.~(\ref{eq:Direqvf})
was previously obtained in Refs.~\cite{CarFul97,SorZil07}. However,
the vector contribution, $\frac{\mathrm{i}}{2}\bar{\gamma}^{a}\gamma_{abc}\eta^{bc}$,
is omitted in these works. For example, in Ref.~\cite{SorZil07},
the incorrect analogue of Eq.~(\ref{eq:3Dm}), in which only the
term $-\mathrm{i}\varepsilon^{dabc}\bar{\gamma}_{d}\bar{\gamma}^{5}$
was accounted for, is used. The correct form of the Dirac equation in a locally Minkowskian frame,
coinciding with Eq.~(\ref{eq:Direqvf}), is derived
in Ref.~\cite{Gas13}.

The vector contribution of the gravitational interaction acts as the
effective electromagnetic field $q_{\mathrm{eff}}A_{\mathrm{eff}}^{a}=-\frac{\mathrm{i}}{2}\gamma^{abc}\eta_{bc}$.
However, the vector potential of this electromagnetic field is imaginary.
It makes the Hamiltonian of the Dirac Eq.~(\ref{eq:Direqvf}) non-Hermitian.
The problem of the non-Hermicity of Hamiltonians of fermions in curved
spacetime was discussed earlier in Refs.~\cite{HuaPar09,GorNez10}.
For instance, following the approach of Ref.~\cite{Ben07}, a nonunitary
transformation of the wave function, which recovers the Hermicity
of the Hamiltonian, was proposed in Ref.~\cite{GorNez10}.

Thus, to develop the approach for the description of the neutrino evolution
in external fields in curved spacetime based on Eq.~(\ref{eq:Direqvftr})
one has to choose the vierbein vectors satisfying the condition $\gamma_{abc}\eta^{bc}=e_{a\ ;\mu}^{\ \mu}=0$.
We assume that this condition is fulfilled.
Thus, we conclude that the neutrino bispinor obeys the following
wave equation in general external fields:
\begin{equation}\label{eq:Direqvftr}
  \left[
    \mathrm{i}\bar{\gamma}^{a}\partial_{a}+
    \bar{\gamma}^{a}\bar{\gamma}^{5}\frac{1}{4}\varepsilon_{abcd}\gamma^{cbd}-
    \frac{\mu}{2}f_{ab}\bar{\sigma}^{ab}-\frac{v^{a}}{2}\bar{\gamma}_{a}+
    \frac{v_{5}^{a}}{2}\bar{\gamma}_{a}\bar{\gamma}^{5}-m
  \right]
  \psi=0,
\end{equation}
where
\begin{equation}\label{eq:V5}
  v_{5}^{a}=v^{a}+\frac{1}{2}\varepsilon^{abcd}\gamma_{cbd},
\end{equation}
is the effective axial-vector field.

The covariant equation for the quasiclassical evolution of the neutrino
spin $s^{a}$ in general external fields in Minkowski space is derived
in Ref.~\cite{DvoStu02}. That derivation of the equation for $s^a$ is based on the Heisenberg equation for the corresponding spin operator accounting for the external fields, which then is averaged over the neutrino wave packet. Using Eqs.~(\ref{eq:Direqvftr}) and~(\ref{eq:V5}),
as well as applying the formalism of Ref.~\cite{DvoStu02}, one gets
that $s^{a}$ obeys the equation,
\begin{equation}\label{eq:sevol}
  \frac{\mathrm{d}s^{a}}{\mathrm{d}\tau}=2\mu
  \left(
    f^{ab}s_{b}-u^{a}f^{bc}u_{b}s_{c}
  \right)+
  \varepsilon^{abcd}v_{b}u_{c}s_{d}+G^{ab}s_{b},
\end{equation}
where
\begin{equation}\label{eq:Gab}
  G^{ab}=
  \left(
    \gamma^{abc}+\gamma^{cab}+\gamma^{bca}
  \right)
  u_{c},
\end{equation}
is the antisymmetric tensor, $G^{ab}=-G^{ba}$, which incorporates
the influence of the gravitational field on the neutrino spin evolution,
$u^{a}$ is the neutrino four velocity, and $\tau$ is the proper
time in the locally Minkowskian frame. Using the properties of $e_{a}^{\ \mu}$, one can show that $\mathrm{d}\tau = \sqrt{g_{\mu\nu}\mathrm{d}x^\mu \mathrm{d}x^\nu}$; i.e., it is invariant under general coordinate transformations.

If we contract $s^a$ with $e_{a}^{\ \mu}$, we get that $S^\mu = s^a e_{a}^{\ \mu}$ transforms as a vector under general coordinate transformations $x^\mu \to x^{\prime \mu}$~\cite{PomKri98}. Thus, basing on the spin vector $s^a$ in a locally Minkowskian frame, we can construct the spin vector $S^\mu$ in an arbitrary frame.

In general situation, the particle four velocity evolves under the
influence of a gravitational field in the locally Minkowskian frame as~\cite{PomKri98}
\begin{equation}\label{eq:uevol}
  \frac{\mathrm{d}u^{a}}{\mathrm{d}\tau}=\tilde{G}^{ab}u_{b},
  \quad
  \tilde{G}^{ab}=\gamma^{abc}u_{c}.
\end{equation}
One can see that Eqs.~(\ref{eq:sevol})-(\ref{eq:uevol}) are inconsistent
with the requirement that $s^{a}u_{a}=0$ in the course of the evolution
of both $s^{a}$ and $u^{a}$. Thus, we have to choose such a vierbein
frame in which $u^{a}=\text{const}$. In this case, Eqs.~(\ref{eq:sevol})
and~(\ref{eq:Gab}) correctly describe the neutrino spin evolution.
It is such a frame, which is taken, e.g., in Refs.~\cite{Dvo06,Dvo13},
where we studied the neutrino circular motion around Schwarzschild
and Kerr BHs.

The first two terms in left-hand side of Eq.~(\ref{eq:sevol}) reproduce
the contributions of the electromagnetic field and the electroweak
interaction with matter to the neutrino spin evolution, when a particle
moves in a curved spacetime, first derived in Ref.~\cite{Dvo13}
based on principles of the general covariance. Now we rederive this
result starting from the more fundamental Dirac Eq.~(\ref{eq:Direq}).
The third term, $G^{ab}s_{b}$, is different from that proposed in Refs.~\cite{PomKri98,Dvo06,Dvo13}:
$\mathrm{d}s^{a}/\mathrm{d}\tau=\tilde{G}^{ab}s_{b}$, where $\tilde{G}^{ab}$
in given in Eq.~(\ref{eq:uevol}). Below, in Secs.~\ref{sec:GW} and~\ref{sec:DISC}, we discuss the discrepancy of the description of the neutrino spin evolution proposed in the present work and that in Refs.~\cite{PomKri98,Dvo06,Dvo13}.

It is convenient to rewrite Eq.~(\ref{eq:sevol}) using the invariant
three vector of the polarization $\bm{\zeta}$, which fixes the neutrino
spin states in the particle rest frame. It is related to $s^a$ by the following expression~\cite{BerLifPit82}:
\begin{equation}\label{eq:sazeta}
  s^a = 
  \left(
    (\bm{\zeta}\mathbf{u}),
    \bm{\zeta} + \frac{\mathbf{u}(\bm{\zeta}\mathbf{u})}{1+u^0}
  \right),
\end{equation}
where $u^{a}=(u^{0},\mathbf{u})$ is the four velocity with respect to the locally Minkowskian coordinates $\bar{x}^a$.
Using the results of
Refs.~\cite{PomKri98,Dvo06,Dvo13} and Eqs.~\eqref{eq:sevol}, \eqref{eq:Gab}, and~\eqref{eq:sazeta}, we get that
\begin{equation}\label{eq:nuspinrot}
  \frac{\mathrm{d}\bm{\zeta}}{\mathrm{d}t}=
  \frac{2}{\gamma}[\bm{\zeta}\times\mathbf{G}],
\end{equation}
where
\begin{equation}\label{eq:vectG}
  \mathbf{G}=\frac{1}{2}
  \left[
    \mathbf{b}_{g}+\frac{(\mathbf{e}_{g}\times\mathbf{u})}{1+u^{0}}
  \right]+
  \frac{1}{2}
  \left[
    \mathbf{u}
    \left(
      v^{0}-\frac{(\mathbf{vu})}{1+u^{0}}
    \right)-
    \mathbf{v}
  \right]+
  \mu
  \left[
    u^{0}\mathbf{b}-\frac{\mathbf{u}(\mathbf{u}\mathbf{b})}{1+u^{0}}+
    (\mathbf{e}\times\mathbf{u})
  \right].
\end{equation}
Here we represent $G_{ab}=(\mathbf{e}_{g},\mathbf{b}_{g})$,
$v^{a}=(v^{0},\mathbf{v})$, $f_{ab}=(\mathbf{e},\mathbf{b})$, $\gamma=U^{0}$,
and $U^{\mu} = (U^0,\mathbf{U})$ is the neutrino four velocity in the world coordinates.

It should be noted that Eqs.~\eqref{eq:nuspinrot} and~\eqref{eq:vectG} are the direct consequence of Eq.~\eqref{eq:sevol}. Thus Eq.~\eqref{eq:nuspinrot} describes the evolution of the invariant spin vector $\bm{\zeta}$. However, we adapt the evolution of $\bm{\zeta}$ for an arbitrary observer having time $t$ in the original world coordinates. That is why the factor $1/\gamma \equiv 1/U^0$ stands in the right-hand side of Eq.~\eqref{eq:nuspinrot}~\cite{PomKri98}.

\section{Neutrino spin oscillations in a gravitational wave\label{sec:GW}}

In this section, we apply the results of Sec.~\ref{sec:FORMAL} to
study neutrino spin oscillations in matter and a magnetic field under
the influence of a plane GW. Some astrophysical applications are also
briefly considered.

Let us take that a plane GW propagates along the $z$ axis.
Choosing the transverse-traceless gauge, we get that the metric has
the form~\cite{Buo07}, 
\begin{equation}\label{eq:metric}
  \mathrm{d}s^{2}=g_{\mu\nu}\mathrm{d}x^{\mu}\mathrm{d}x^{\nu}=
  \mathrm{d}t^{2}-
  \left(
    1-h_{+}\cos\phi
  \right)
  \mathrm{d}x^{2}-
  \left(
    1+h_{+}\cos\phi
  \right)
  \mathrm{d}y^{2}+2\mathrm{d}x\mathrm{d}y h_{\times}\sin\phi-\mathrm{d}z^{2},
\end{equation}
where $h_{+}$ and $h_{\times}$ are the dimensionless amplitudes
of two independent polarizations of the wave, $\phi=\left(\omega t-kz\right)$
is the phase of the wave, $\omega$ is frequency of the wave, and
$k$ is the wave vector. In Eq.~(\ref{eq:metric}), we use Cartesian
world coordinates $x^{\mu}=(t,x,y,z)$. In the following, we 
consider GW with the circular polarization in which $h_{+}=h_{\times}=h$.
To discriminate between left and right polarizations of the wave we
introduce the parameter $\epsilon=\pm1$ in the phase of the wave
$\phi\to\epsilon\left(\omega t-kz\right)$.

Neutrino spin evolution is governed by Eq.~(\ref{eq:nuspinrot}).
To proceed in the study of neutrino spin oscillations we should find all the parameters in Eq.~(\ref{eq:vectG})
in the locally Minkowskian frame. One can check that the following vierbein vectors:
\begin{align}\label{eq:vierbexpl}
  e_{0}^{\ \mu}= &
  \left(
    1,0,0,0
  \right),
  \nonumber
  \\
  e_{1}^{\ \mu}= & \frac{1}{\sqrt{1+h}}
  \left(
    0,-\sin\frac{\phi}{2},\cos\frac{\phi}{2},0
  \right),
  \nonumber
  \\
  e_{2}^{\ \mu}= & \frac{1}{\sqrt{1-h}}
  \left(
    0,\cos\frac{\phi}{2},\sin\frac{\phi}{2},0
  \right),
  \nonumber
  \\
  e_{3}^{\ \mu}= &
  \left(
    0,0,0,1
  \right),
\end{align}
diagonalize the metric in Eq.~(\ref{eq:metric}). Note that Eq.~(\ref{eq:vierbexpl})
exactly accounts for the amplitude of the wave.

First, we have to check the Hermicity of the Hamiltonian of Eq.~(\ref{eq:Direqvf}).
Using Eqs.~(\ref{eq:metric}) and~(\ref{eq:vierbexpl}), one gets
that $e_{a\ ;\mu}^{\ \mu}=0$. Thus Eqs.~(\ref{eq:Direqvf}) and~(\ref{eq:Direqvftr})
coincide.

We consider the situation when neutrinos are emitted by the same source
of GWs. It is a reasonable situation when we study oscillations of
astrophysical neutrinos. In this case, $\mathbf{U}=\mathrm{d}\mathbf{x}/\mathrm{d}s=(0,0,U_{z})$.
Using Eq.~(\ref{eq:uevol}), one can show that the acceleration of
such neutrinos in the locally Minkowskian frame is absent: $\tfrac{\mathrm{d}u^{a}}{\mathrm{d}\tau}=\tilde{G}^{ab}u_{b}\equiv0$.
It means that the change of the polarization $(\bm{\zeta}\mathbf{u})$
is entirely defined by the neutrino spin evolution in Eq.~(\ref{eq:nuspinrot}).
It should be also noted that $u_{z}=U_{z}=\text{const}$ and $u^{0}=U^{0}=\text{const}$.

Moreover we suppose that, besides GW, a neutrino interacts with nonmoving
and unpolarized matter, i.e. $V^{0}\neq0$ and $\mathbf{V}=0$ in
Eq.~(\ref{eq:Vmudef}). We also take that a constant uniform magnetic
field transverse to the neutrino motion is present in the world coordinates
$x^{\mu}$. For example, we suppose that $\mathbf{B}=(B,0,0)$. In
this situation, $v^{0}=V^{0}$, 
\begin{equation}
  b_{x}=\frac{B}{\sqrt{1-h}}\sin\frac{\phi}{2},
  \quad
  b_{y}=-\frac{B}{\sqrt{1+h}}\cos\frac{\phi}{2},
\end{equation}
and $\mathbf{e}=0$.

Now we can find the components of the vector $\bm{\Omega}=\mathbf{G}/\gamma$,
which defines the neutrino spin evolution, as
\begin{align}\label{eq:Omegalin}
  \Omega_{x}= & \frac{\mu B}{\sqrt{1-h}}\sin\frac{\phi}{2},
  \nonumber
  \\
  \Omega_{y}= & -\frac{\mu B}{\sqrt{1+h}}\cos\frac{\phi}{2},
  \nonumber
  \\
  \Omega_{z}= & \frac{V^{0}U_{z}}{2U^{0}}+
  \frac{\epsilon(kU_{z}-\omega U^{0})}{4U^{0}\sqrt{1-h^{2}}}.
\end{align}
Using Eqs.(\ref{eq:nuspinrot}), (\ref{eq:vectG}), and~(\ref{eq:Omegalin}),
the evolution of the neutrino polarization can be rewritten using
the effective neutrino wave function $\nu^{\mathrm{T}}=(\nu_{+},\nu_{-})$,
which obeys the Schr\"odinger equation~\cite{LobPav99,EgoLobStu00},
\begin{equation}\label{eq:effSchrod}
  \mathrm{i}\frac{\mathrm{d}\nu}{\mathrm{d}t}=H_{\mathrm{eff}}\nu,
  \quad
  H_{\mathrm{eff}}=-(\bm{\sigma}\cdot\bm{\Omega}),
\end{equation}
where $\bm{\sigma}=\left(\sigma_{1},\sigma_{2},\sigma_{3}\right)$
are the Pauli matrices. Here $\pm$ mark different neutrino helicities, i.e. the projections of the invariant three vector of the neutrino spin on the neutrino velocity, $(\bm{\zeta} \mathbf{u})/|\mathbf{u}| = \pm 1$. 

We mentioned in Sec.~\ref{sec:FORMAL} that the different contribution of the gravity to the neutrino spin evolution equation was obtained in Refs.~\cite{Dvo06,Dvo13} using the equivalence principle, namely, $\mathrm{d}s^{a}/\mathrm{d}\tau=\tilde{G}^{ab}s_{b}$, where $\tilde{G}^{ab}$ is given in Eq.~(\ref{eq:uevol}). One may expect that the equation and Eqs.~\eqref{eq:sevol} and~\eqref{eq:Gab} can result in distinct descriptions of neutrino spin oscillations. To study this issue we have used the approach in Refs.~\cite{Dvo06,Dvo13} to obtain the analogue of the effective Hamiltonian in Eq.~\eqref{eq:effSchrod} for a neutrino interacting with matter, the transverse magnetic field and the plane GW, with a neutrino propagating along the wave. It turns out to coincide with that in Eqs.~\eqref{eq:Omegalin} and~\eqref{eq:effSchrod}. It means that the approach developed here and that in Refs.~\cite{PomKri98,Dvo06,Dvo13} are equivalent, at least for the particular physics system. Since these two methods give coinciding results, we omit detailed calculations based on the $\mathrm{d}s^{a}/\mathrm{d}\tau=\tilde{G}^{ab}s_{b}$ equation.

It is convenient to modify the effective wave function as $\nu=\exp\left[-\mathrm{i}\sigma_{3}\left(\dot{\phi}t+\pi\right)/4\right]\tilde{\nu}$,
where $\dot{\phi}=\epsilon\left(\omega-kU_{z}/U^{0}\right)$. Taking
into account that $h\ll1$, we get that
%
%
\begin{equation}\label{eq:effSchrodtilde}
  \mathrm{i}\frac{\mathrm{d}\tilde{\nu}}{\mathrm{d}t}=
  \tilde{H}_{\mathrm{eff}}\tilde{\nu},
  \quad
  \tilde{H}_{\mathrm{eff}}=
  \left(
    \begin{array}{cc}
      -V^{0}/2 & \mu B
      \left[
        1-he^{-\mathrm{i}\dot{\phi}t}/2
      \right]
      \\
      \mu B
      \left[
        1-he^{\mathrm{i}\dot{\phi}t}/2
      \right] & V^{0}/2
    \end{array}
  \right).
\end{equation}
In Eq.~\eqref{eq:effSchrodtilde}, we assume that neutrinos are ultrarelativistic,
i.e. $U_{z}=\beta U^{0}$, where $\beta \approx 1$ is the neutrino
velocity.

Now it is interesting to compare neutrino spin oscillations in a plane
electromagnetic wave, studied in Refs.~\cite{EgoLobStu00,Dvo18,Dvo19},
with oscillations in a plane GW with the circular polarization. Unlike
an electromagnetic wave, GW cannot induce a neutrino spin flip. Indeed, if $B=0$,
there are no transitions $\nu_{-}\leftrightarrow\nu_{+}$.
This result coincides with the finding of Ref.~\cite{Qua16}.

GW can only influence the resonance condition in neutrino oscillations.
Indeed, choosing $\dot{\phi}$ in Eq.~(\ref{eq:effSchrodtilde})
in a certain way, we can reach a significant enhancement of the probability
of neutrino spin oscillations $\nu_{-}\to\nu_{+}$,
$P_{\nu_{-}\to\nu_{+}}(t)\equiv|\nu_{+}|^{2}=|\tilde{\nu}_{+}|^{2}$,
calculated using the solution of Eq.~(\ref{eq:effSchrodtilde}). 
Such an enhancement of the transition probability is known as the parametric resonance in neutrino oscillations.

We assume that initially all neutrinos are left polarized, i.e.
$\nu^\mathrm{T}(0) = (0, 1)$, as it should be if neutrinos are created in electroweak processes.
Of course, the exact initial condition should be $\nu^\mathrm{T}_\text{exact}(0) = (\sqrt{P_+},\sqrt{P_-})$, where $P_\pm = (1 \mp \beta)/2$. However, using $\nu_\text{exact}(0)$ practically does not change the numerical solution of Eq.~\eqref{eq:effSchrodtilde} in the case of relativistic particles.

For the first time, the parametric resonance in neutrino oscillations
in matter with harmonically varying density was studied in Ref.~\cite{Akh88}.
One can see in Eq.~(\ref{eq:effSchrodtilde}) that the action of
GW on neutrino oscillations is the effective harmonic modulation of
the transverse magnetic field. It is this manifestation of the parametric
resonance, which was discussed in Ref.~\cite{DvoStu04}, where neutrino
spin and spin-flavor oscillations in inhomogeneous electromagnetic
fields were studied.

Using the results of Ref.~\cite{DvoStu04}, we suppose that 
\begin{equation}\label{eq:rescond}
  \dot{\phi}=2\Omega,
\end{equation}
where $\Omega=\sqrt{(\mu B)^{2}+V_{0}^{2}/4}$ is the frequency of
the neutrino spin precession at the absence of GW, $\dot{\phi} = \omega m^2/2 E^2$ for relativistic neutrinos, and $E$ is the neutrino energy. 
Here we neglect the dispersion of GW and set $\omega=k$.
If $h\ll1$, Eq.~(\ref{eq:effSchrodtilde})
can be reduced to the Mathieu equation. The solution of such an equation
can be represented in terms of special functions only. That is why
we analyze here only numerical solutions of Eq.~(\ref{eq:effSchrodtilde})
to highlight the manifestation of the parametric resonance.

We suppose that a beam of left polarized neutrinos is produced at
the distance $z_{0}=10^{-5}\,\text{au}$ from merging BHs. Here $1\,\text{au}=1.5\times10^{13}\,\text{cm}$
is the astronomical unit. There is no commonly adopted model for the production of neutrinos in coalescing BHs or NSs. However, neutrinos can be produced in dense matter of an accretion disk surrounding these BHs. If BHs have masses $\sim 30 M_\odot$~\cite{Abb16}, their Schwartzchild radii are $R_\mathrm{S} \approx 10^7\,\text{cm}$. The inner radius of such a disk was found in Ref.~\cite{Kha18} to be $\sim 10 R_\mathrm{S} = 10^8\,\text{cm}$, which is comparable with $z_0$.

The considered BHs are taken to emit GW in the same
direction as the neutrino beam. We assume that, at $z_{0}=10^{-5}\,\text{au}$,
$h=10^{-1}$. If we observed the merging of these BHs at the distance
$z\sim1\,\text{Gpc}$~\cite{Abb18}, the amplitude of the produced
GW would be $h_{\text{obs}}=5\times10^{-21}$, which is very close
to values recorded by the modern GW detectors~\cite{Abb16}.

We assume that the accretion disk, which surrounds coalescing BHs and where neutrinos are produced, consists of the electroneutral hydrogen plasma
with the electron number density in the range $n_{e}=(10^{18}-10^{22})\,\text{cm}^{-3}$.
In this case, the effective potential of the neutrino interaction with
matter reads $V^{0}=\sqrt{2}G_{\mathrm{F}}n_{e}$. Note that numerical simulations of accretion disks around merging BHs predict the maximal plasma densities $\sim (10^{29} - 10^{30})\,\text{cm}^{-3}$, which is much higher than the value chosen here. However, since we  assume that the plasma density is constant along the neutrino trajectory, we may use these moderate values.

We also suppose that a constant magnetic field  is present in the system. The magnetic energy is taken to be $\mu B \approx 5.8\times (10^{-20} - 10^{-16})\,\text{eV}$. If we suppose that $\mu=10^{-14}\mu_{\mathrm{B}}$~\cite{10-14}, where
$\mu_{\mathrm{B}}$ is the Bohr magneton, then the magnetic field should be in the range $(10^3 - 10^7)\,\text{G}$. Even if we take that $\mu \sim 10^{-19}\mu_{\mathrm{B}}$~\cite{FujShr80}, which is a natural value for a Dirac neutrino with the mass $\sim 1\,\text{eV}$~\cite{Ase11}, the magnetic field should be as strong as $(10^8 - 10^{12})\,\text{G}$. Such magnetic fields are compatible with the results of the numerical simulations of accretion disks carried out in Ref.~\cite{Kha18}, where $B\sim10^{12}\,\text{G}$ was used.

Neutrinos are taken to have the mass $m\sim1\,\text{eV}$~\cite{Ase11}
and the energy in the range $E=(10 - 10^3)\,\text{eV}$.
To fulfil the resonance condition in Eq.~(\ref{eq:rescond}) for
the chosen parameters of neutrinos and the external fields, we have to
set $\omega=5.4\times10^{2}\,\text{s}^{-1}$. If one observed GWs,
emitted by the merging of BHs, from the distance $z\sim1\,\text{Gpc}$~\cite{Abb18},
the GW frequency would be red-shifted to $\omega_{\text{obs}}=\omega/(1+\mathrm{z})=4.5\times10^{2}\,\text{s}^{-1}$,
where $\mathrm{z}\approx0.2$~\cite{Abb17}. This value of $\omega_{\text{obs}}$
is again very close to GW frequencies registered by the current GW detectors~\cite{Abb16}.

If one takes the parameters of the external fields and a neutrino as specified above, the resonance condition in Eq.~\eqref{eq:rescond} is fulfiled. Indeed, in the numerical simulations of Eq.~\eqref{eq:effSchrodtilde}, we use mainly two sets of parameters: $n_e = 10^{22}\,\text{cm}^{-3}$, $\mu B = 5.8\times 10^{-16}\,\text{eV}$, and $E = 10\,\text{eV}$; or $n_e = 10^{18}\,\text{cm}^{-3}$, $\mu B = 5.8\times 10^{-20}\,\text{eV}$, and $E = 10^3\,\text{eV}$. In both cases we take that $m = 1\,\text{eV}$ and $\omega = 5.4\times10^{2}\,\text{s}^{-1}$. Now, if one computes $\dot{\phi}$ and $\Omega$, one can see that $\dot{\phi}=2\Omega$, i.e. Eq.~\eqref{eq:rescond} is satisfied. It should be noted that the magnetic field $B$ and the electron density $n_e$ are assumed to be constant along the neutrino trajectory. This assumption allows one to simplify the analysis of Eq.~\eqref{eq:effSchrodtilde}. Moreover, in  this situation, we can highlight the manifestation of the parametric resonance in neutrino oscillations.

\begin{figure}
  \centering
  \subfigure[]
  {\label{1a}
  \includegraphics[scale=.45]{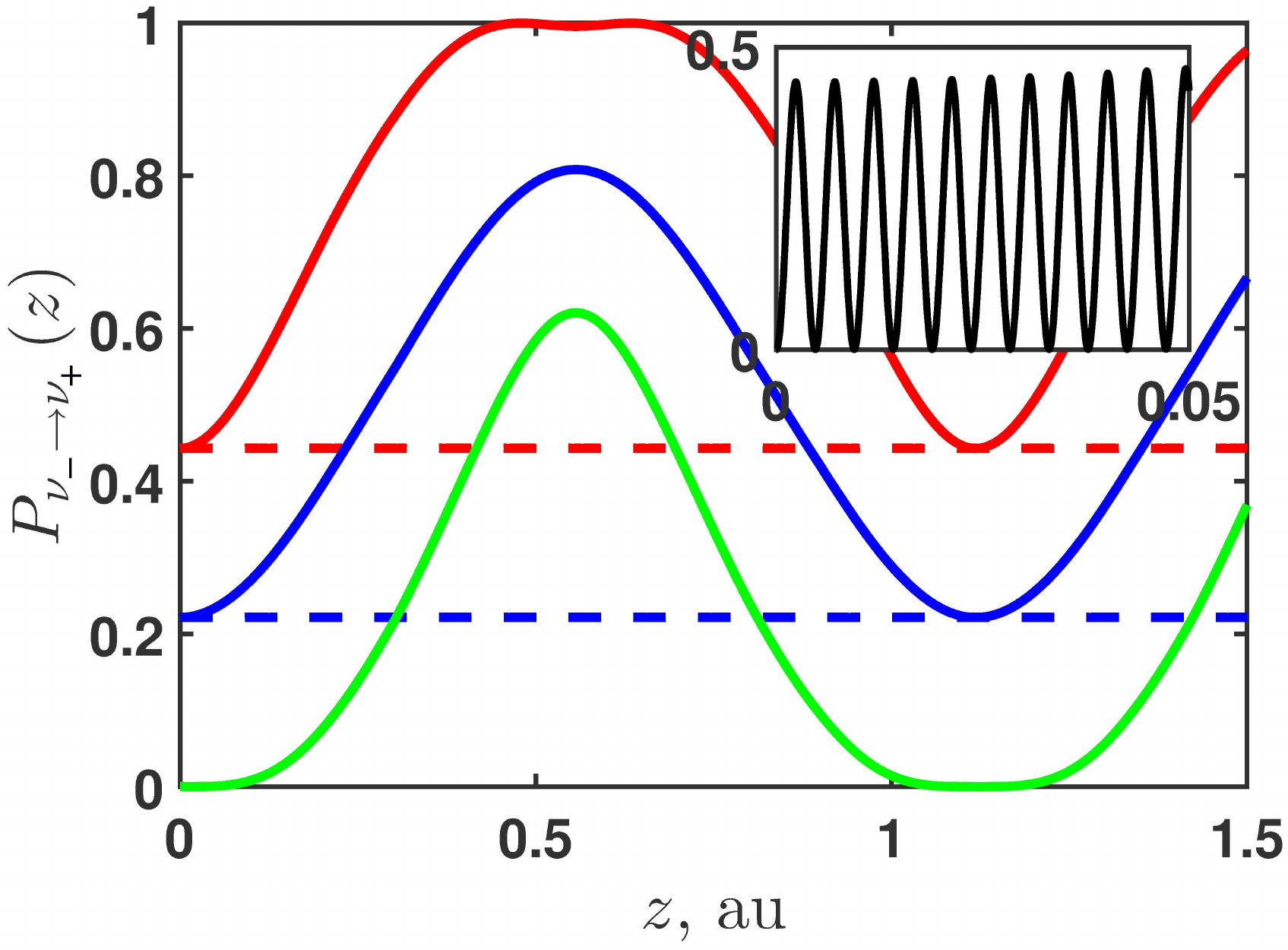}}
  \hskip-.2cm
  \subfigure[]
  {\label{1b}
  \includegraphics[scale=.45]{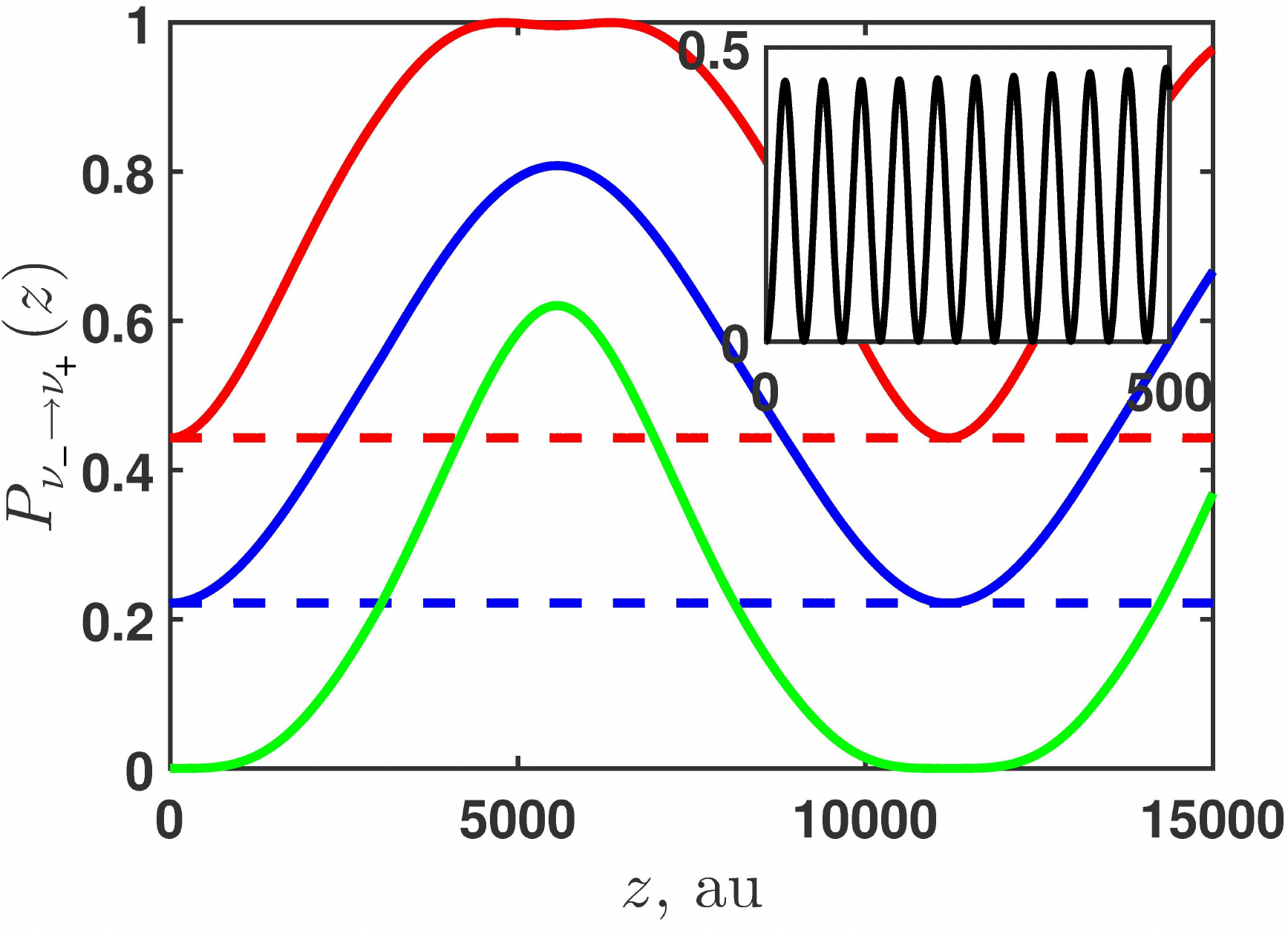}}
  \protect
  \caption{The probabilities $P_{\nu_{-}\to\nu_{+}}(z)$ for transitions
  $\nu_{-}\to\nu_{+}$ versus the distance $z=t$,
  passed by the neutrino beam, built on the basis of the numerical solution
  of Eq.~(\ref{eq:effSchrodtilde}) for different plasma densities, magnetic fields
  and neutrino energies.
  Red and green lines are the upper
  and lower envelope functions. Blue lines are the averaged transition
  probabilities. Solid lines correspond to $h=10^{-1}$ and dashed lines
  to $h=0$.
  (a) $n_e = 10^{22}\,\text{cm}^{-3}$, $\mu B = 5.8\times 10^{-16}\,\text{eV}$,
  and $E = 10\,\text{eV}$;
  (b) $n_e = 10^{18}\,\text{cm}^{-3}$, $\mu B = 5.8\times 10^{-20}\,\text{eV}$,
  and $E = 10^3\,\text{eV}$.
  The insets represent the transition probabilities, shown
  by the black lines, for $0<z<5\times10^{-2}\,\text{au}$ in panel~(a)
  and $0<z<5\times10^{2}\,\text{au}$ in panel~(b).
  \label{fig:PLR}}
\end{figure}

In Fig.~\ref{fig:PLR}, we show the transition probabilities $P_{\nu_{-}\to\nu_{+}}(z)$
versus the distance traveled by the neutrino beam $z\approx t$ for
spin oscillations $\nu_{-}\to\nu_{+}$ obtained
using the numerical solution of Eq.~(\ref{eq:effSchrodtilde}) for different plasma densities, magnetic fields and neutrino energies chosen above. The
function $P_{\nu_{-}\to\nu_{+}}(z)$ is rapidly oscillating.
That is why we represent it only in the insets in Fig.~\ref{fig:PLR}
for small $z$. In the main Fig.~\ref{fig:PLR},
we show only the upper and lower envelope functions, which are built
with the help of the spline interpolation of the maxima and minima of
$P_{\nu_{-}\to\nu_{+}}(z)$, respectively, as well as the averaged
transition probabilities $\bar{P}_{\nu_{-}\to\nu_{+}}$. To
highlight the impact of GW on neutrino spin oscillations, we show
the upper envelope functions of $P_{\nu_{-}\to\nu_{+}}(z)$,
as well as $\bar{P}_{\nu_{-}\to\nu_{+}}$, at $h=0$, i.e., for
neutrino spin oscillations only in matter and the transverse magnetic
field.

One can see in Fig.~\ref{fig:PLR} that, at $z\approx0.5\,\text{au}$ for $E = 10\,\text{eV}$ and at $z\approx 500\,\text{au}$ for $E = 10^3\,\text{eV}$,
the upper envelope functions reach the unit value and $\bar{P}_{\nu_{-}\to\nu_{+}}\sim0.75$.
It is the manifestation of the parametric resonance, which happens
even for small $h=10^{-1}$. The upper envelope functions and $\bar{P}_{\nu_{-}\to\nu_{+}}$
for $h=0$ do not exceed $0.5$ and $0.25$, respectively. It means
that the transition probability cannot be amplified to great values
without GW.

If the resonance condition in Eq.~\eqref{eq:rescond} is satisfied, the amplitude of the solution of the Mathieu equation starts to grow as $\sim \exp(\varkappa t)$~\cite{BogMit57}, where the increment reads $\varkappa \sim h \dot{\phi}$. Thus, if a neutrino travels the distance $L = \varkappa^{-1} \sim (h \dot{\phi})^{-1}$, the transition probability reaches great values. Comparing Figs.~\ref{1a} and~\ref{1b}, we can confirm this feature.

If we study spin oscillations of low energy neutrinos with $E = 10\,\text{eV}$ shown in Fig.~\ref{1a}, the value of $z_\mathrm{max} = L \sim 0.5\,\text{au}$ is comparable with the size of a planetary system. It means that spin oscillations induced by GW are important for such neutrinos. Note that the creation of low energy Dirac neutrinos with $E \lesssim 10\,\text{eV}$ in matter with a time dependent density was studied in Ref.~\cite{DvoGavGit14}.

The outer radius of an accretion disk for BH with the mass $10 M_\odot$ was found in Ref.~\cite{Oku05} to reach $5 \times 10^{4} R_\mathrm{S} = 5\times 10^{11}\,\text{cm}$, which is only 1 order of magnitude less than $z_\mathrm{max}$ for  $E = 10\,\text{eV}$. However, if we deal with two merging BHs with $M = 30 M_\odot$~\cite{Abb16}, they can attract some additional matter from the outer space slightly increasing the size of the mutual accretion disk.

\begin{figure}
  \centering
  \subfigure[]
  {\label{2a}
  \includegraphics[scale=.45]{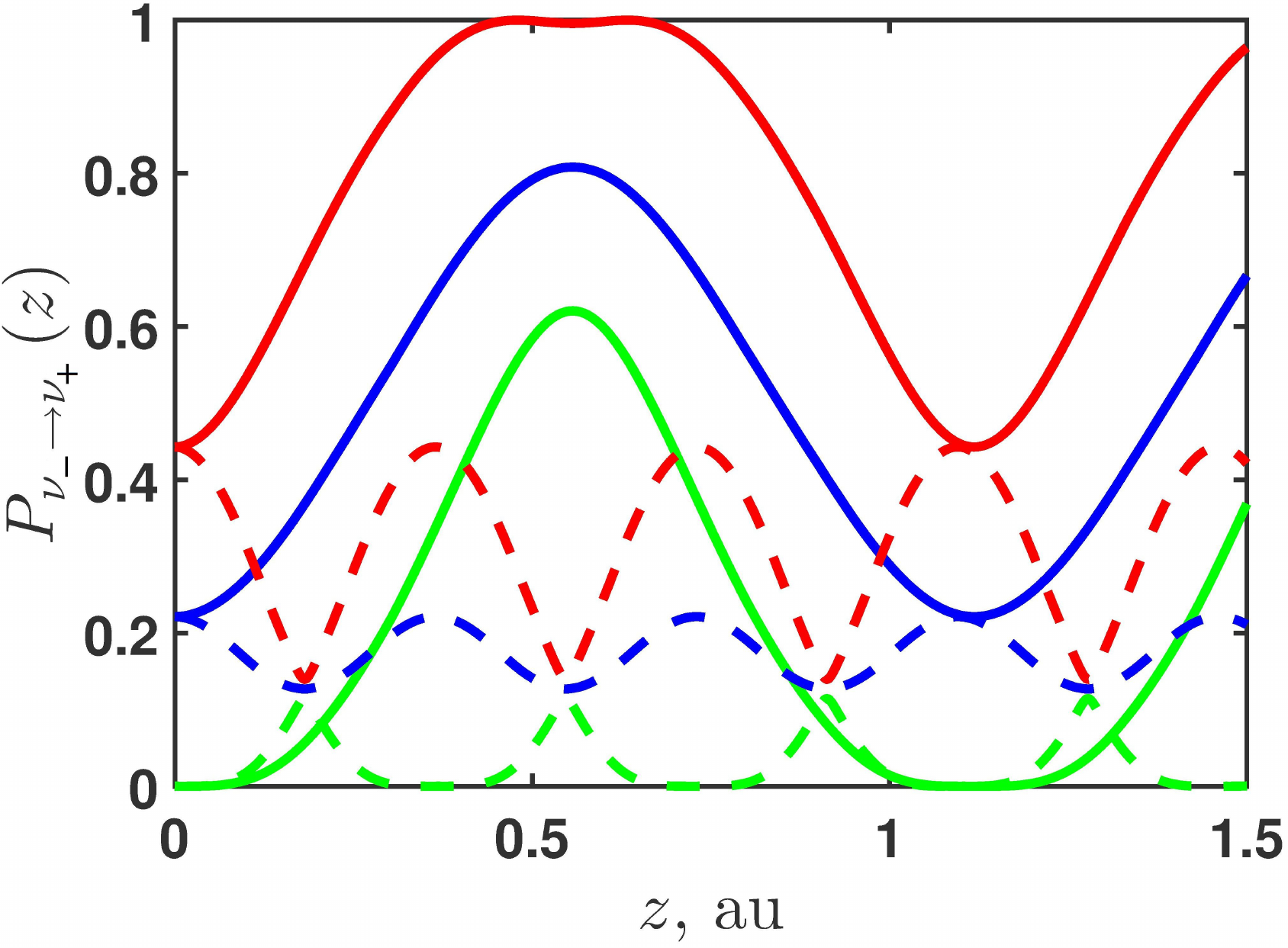}}
  \hskip-.2cm
  \subfigure[]
  {\label{2b}
  \includegraphics[scale=.45]{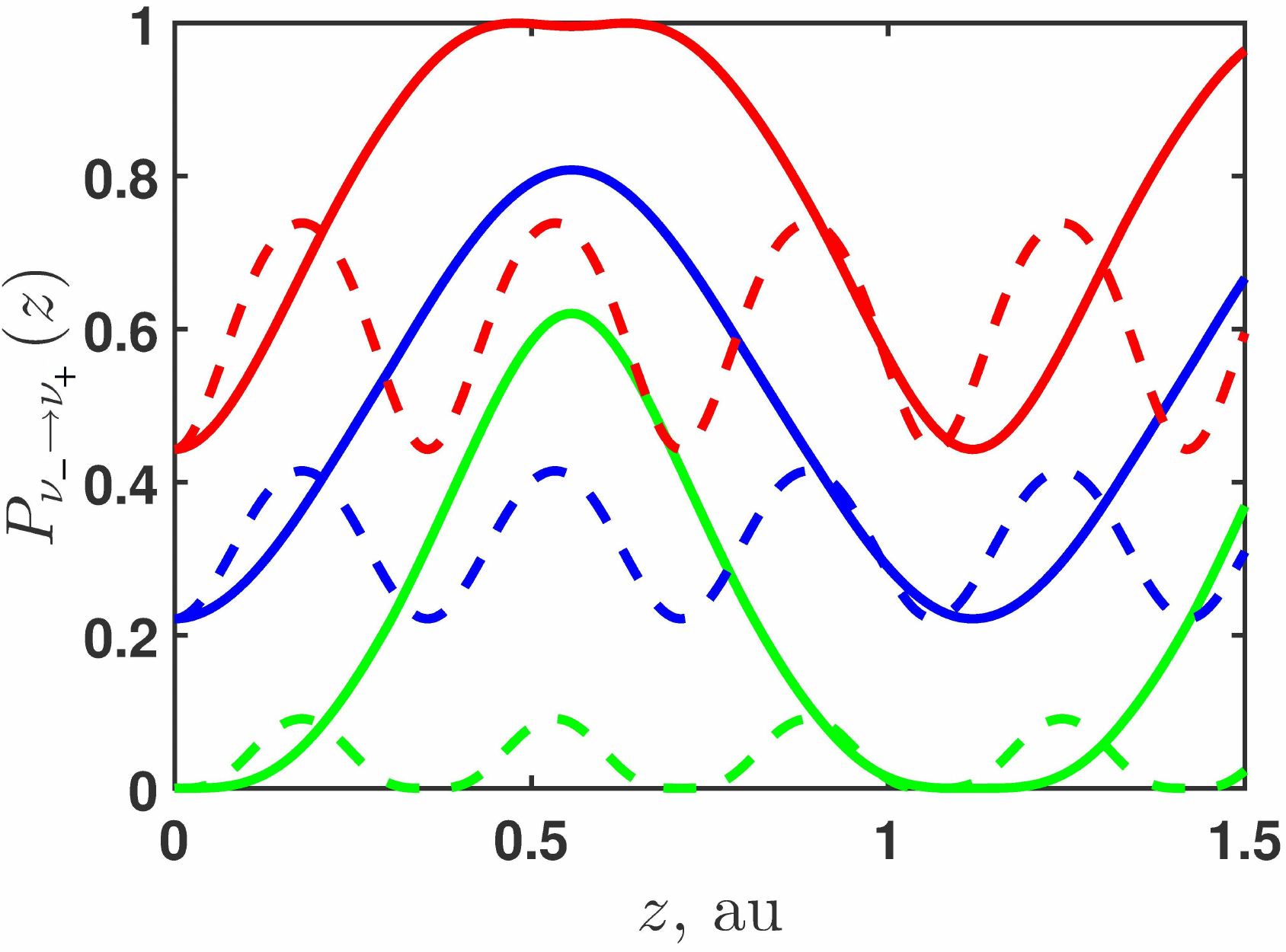}}
  \protect
  \caption{The behavior of neutrino spin oscillations for different $\dot{\phi}$.
  The upper (red lines) and lower (green lines) envelope functions for
  $P_{\nu_{-}\to\nu_{+}}(z)$, as well as the averaged transition
  probabilities (blue lines), versus the distance traveled by the neutrino
  beam built on the basis of the numerical solution of Eq.~(\ref{eq:effSchrodtilde}).
  Solid lines imply the resonance condition in Eq.~(\ref{eq:rescond})
  and correspond to the same parameters as in Fig.~\ref{1a}.
  Dashed lines correspond to small deviations $\dot{\phi}_\pm$ from Eq.~(\ref{eq:rescond}).
  (a) $\dot{\phi}_+=2\Omega+h\Omega/4$. (b) $\dot{\phi}_-=2\Omega-h\Omega/4$.   
  \label{fig:resconddev}}
\end{figure}

Note that nowadays there are active searches of simultaneous emission of GWs and neutrinos in coalescence of NSs~\cite{Alb19}. If we suppose that the emission of low energy neutrinos happens owing to the time dependence of the effective number density of NS, as in Ref.~\cite{DvoGavGit14}, the duration of the neutrino pulse is $\sim 10^{-3}\,\text{s}$. According to up-to-date observations~\cite{Abb18}, a typical time scale of a GW wave packet is $\sim 10^{-1}\,\text{s}$. Thus, these neutrinos are very well localized inside such a wave packet. Thus, in the first approximation, in our studies of neutrino oscillations driven by GW, we can neglect the time dependence of the frequency and the amplitude of GW (the GW chirp), observed by the modern GW detectors~\cite{Abb18}.

The condition of the parametric resonance appearance in Eq.~(\ref{eq:rescond})
is very sensitive to the change of $\dot{\phi}$ or, equivalently,
to different frequencies $\omega$ of GW. To illustrate the dependence
of the dynamics of neutrino oscillations for different $\dot{\phi}$,
in Fig.~\ref{fig:resconddev}, we show the upper and lower envelope
functions, as well as the averaged transition probabilities, versus
the distance passed by the neutrino beam for different $\dot{\phi}$. We compare the case when
the resonance condition in Eq.~(\ref{eq:rescond}) is fulfiled with
the small deviations from this condition, $\dot{\phi}_\pm=2\Omega\pm h\Omega/4$.
One can see in Fig.~\ref{fig:resconddev} that in both cases the
upper envelope function and $\bar{P}_{\nu_{-}\to\nu_{+}}$ for
$\dot{\phi}\neq2\Omega$ have smaller values than these for $\dot{\phi}=2\Omega$.
Moreover, comparing Figs.~\ref{2a} and~\ref{2b},
one gets that the transition probability behaves differently in
decreasing and increasing of $\omega$.

Now let us qualitatively discuss the dynamics of neutrino oscillations at larger distances from the point of the BHs merging. We suppose that, in the vicinity of this point, the resonance condition in Eq.~\eqref{eq:rescond} is fulfiled. It means that the quantities $\dot{\phi}$, $\mu B$, and $V_0$ are of the same order of magnitude there. If a neutrino propagates further, the magnetic field decreases as $1/r^3$ i.e. quite rapidly. The region of the BH coalescence, where neutrinos are produced, is likely to be surrounded by dense matter, e.g., an accretion disk. However, the matter density also should decrease very rapidly outside an accretion disk at great $r$. The frequency of GW, which $\dot{\phi}$ is proportional to, is practically unchanged. It means that, if Eq.~\eqref{eq:rescond} is valid at moderate $r$, at great $r$, it is converted to
\begin{equation}\label{eq:greatdist}
  \dot{\phi}\gg\Omega.
\end{equation}
We should describe neutrino spin oscillations if the condition in Eq.~\eqref{eq:greatdist} is fulfiled.

Of course, to study neutrino spin oscillations in this case we can employ numerical simulations for $r_0 < z < r$, where $r \gg r_0$. However, at great $r$, we have to take into account the fact that GW become spherical rather than plane, as we assumed in Eq.~\eqref{eq:metric}. That is why we qualitatively describe spin oscillations in this situation.

The analysis of Eq.~\eqref{eq:effSchrodtilde} accounting for Eq.~\eqref{eq:greatdist} at arbitrary $h$ was carried out in Ref.~\cite{Dvo07}. In this situation, the effective Hamiltonian has the constant term and the rapidly oscillating one with the frequency $\dot{\phi}$. We split the wave function into slowly and rapidly oscillating parts, $\tilde{\nu} = \langle \tilde{\nu} \rangle + \delta \tilde{\nu}$. After averaging over $2\pi/\dot{\phi}$, we find that $\langle \tilde{\nu} \rangle$ obeys the equation~\cite{Dvo07},
\begin{equation}\label{eq:Schrrapid}
  \mathrm{i}\frac{\mathrm{d} \langle \tilde{\nu} \rangle}{\mathrm{d}t} =
  \tilde{H}_{\mathrm{eff}} \langle \tilde{\nu}\rangle,
  \quad
  \tilde{H}_{\mathrm{eff}} =
  \left(
    \begin{array}{cc}
      -V^{0}/2 - (\mu B h)^2/4\dot{\phi} & \mu B\\
      \mu B & V^{0}/2 + (\mu B h)^2/4\dot{\phi}
    \end{array}
  \right),
\end{equation}
where we take that $\epsilon = -1$ in $\dot{\phi}$.

Using Eq.~\eqref{eq:Schrrapid} and supposing that external fields, $V^0$, $B$, and $h$, change adiabatically, we get that the resonance condition reads
\begin{equation}\label{eq:rescondrapid}
  V^{0} = \frac{(\mu B h)^2}{2\omega(1-\beta)}.
\end{equation}
There is a very poor knowledge on the matter density distribution, which $V^0$ is proportional to, in an accretion disk. This quantity is rather model dependent. However, let us we assume that, in the vicinity of the point of BH merging, the matter term $V^0$ dominates over other contributions in $\tilde{H}_{\mathrm{eff}}$ in Eq.~\eqref{eq:effSchrodtilde}. An analogous assumption was made to build Fig.~\ref{fig:PLR}. Then we take that an accretion disk has a sharp edge at $r_\mathrm{out}$. Thus, the condition in Eq.~\eqref{eq:rescondrapid} can be fulfiled near $r_\mathrm{out}$ since both $B$ and $h$ are nonzero even outside the accretion disk.

\section{Discussion\label{sec:DISC}}

In the present work, we have studied the neutrino spin evolution in
background matter and an external electromagnetic field in curved
spacetime. This study was motivated by the necessity for the substantiation
of the quasiclassical equation for the neutrino spin evolution, which
was proposed in Ref.~\cite{Dvo13}. In Sec.~\ref{sec:FORMAL}, we
have rederived this covariant equation starting from the Dirac equation
for a massive neutrino interacting with external fields in curved
spacetime.

The form of the tensor $G_{ab}$ in Eq.~(\ref{eq:Gab}), which incorporates
the contribution of the gravitational interaction to the neutrino
spin evolution, is different from that found in Refs.~\cite{PomKri98,Dvo06,Dvo13}.
We have compared the phenomenological consequences, namely, the description of neutrino spin oscillations in particular background fields in curved spacetime, of the approach in Refs.~\cite{PomKri98,Dvo06,Dvo13} and that developed here. They turn out to coincide. It is likely to be a general result. Indeed, as shown in Ref.~\cite{SorZil07}, these approaches predict the same frequency of the neutrino spin precession when a particle circularly orbits a Schwarzschild BH. However, the problem of the most precise description of the particle spin evolution in background fields and an arbitrary gravitational field is still under discussion (see, e.g., the series of works~\cite{ObuSilTer09,ObuSilTer17}). Nevertheless, we should mention that the contributions of background matter and
an electromagnetic field to the neutrino spin evolution Eq.~(\ref{eq:sevol})
coincide with those proposed in Ref.~\cite{Dvo13}.

Then, in Sec.~\ref{sec:GW}, we have applied the developed formalism
for the description of neutrino spin oscillations in matter and the
constant transverse magnetic field under the influence of a plane
GW. We have derived the effective Schr\"odinger equation and demonstrated
that GW can induce the parametric resonance in neutrino spin oscillations.

It should be noted that the contribution to the neutrino spin evolution
from the proper gravitational field of a compact rotating object depends
on the distance between a neutrino and this object as $1/r^{3}$~\cite{Dvo06}.
The amplitude of GW falls with the distance as $1/r$. Hence, despite
the fact that the impact of the proper gravitational interaction on the spin precession
is stronger than that of GW at short distances, neutrino spin oscillations
can be more significantly affected by GW at larger distances where
the direct gravitational interaction decays more rapidly.

Considering GW emitted by coalescing BHs, in Sec.~\ref{sec:GW}, we have demonstrated that
the transition probability of neutrino spin oscillations can potentially
reach great values. In our simulations we have chosen the parameters of external fields, such as the matter density and the strength of the magnetic field, not exceeding the predictions of the model of accretion disks around BHs~\cite{Kha18}. The range of neutrino magnetic moments, required for the enhancement of the transition probability, is also compatible with the values expected for Dirac neutrinos~\cite{Bel05,FujShr80}.

Unfortunately, in Fig.~\ref{1a}, we predict the oscillations length $L$ comparable with the outer radius of an accretion disk~\cite{Oku05} only for very low energy neutrinos with $E = 10\,\text{eV}$. Nevertheless the creation of such neutrinos is anticipated in matter with time dependent density~\cite{DvoGavGit14}.

In Sec.~\ref{sec:GW}, we have also studied the dynamics of neutrino spin oscillations
for different frequencies of GW; cf., Fig.~\ref{fig:resconddev}. We have demonstrated that small deviations from the resonance condition in Eq.~\eqref{eq:rescond} result in the decreasing of the transition probability.

We have qualitatively studied neutrino spin oscillations at large distances from coalescing BHs at the end of Sec.~\ref{sec:GW}. For this purpose we have assumed that both the magnetic and gravitational fields, as well as the matter density, are rapidly decreasing at large distances. Thus $\dot{\phi}$ becomes the greatest parameter in the effective Hamiltonian. In this situation, using the results of Ref.~\cite{Dvo07}, it was possible to exclude the time dependence from the effective Hamiltonian and analyze whether a resonance could appear in neutrino oscillations. One could expect the resonant enhancement of spin oscillations if an accretion disk has a sharp outer edge. However, the actual situation is unclear because of the lack of understanding of accretion disks dynamics around BHs.



\begin{acknowledgments}
I am thankful to A.~Buonanno, V.~I.~Dokuchaev, M.~Gasperini, Yu.~N.~Obukhov,
K.~A.~Postnov, O.~V.~Teryaev, and A.~F.~Zakharov for useful
comments. This work was partially supported by RFBR (Grant No. 18-02-00149a).
\end{acknowledgments}


\begin{thebibliography}{50}
\bibitem{10-14}
  The chosen value of $\mu$ does not exceed the theoretical upper bound on magnetic moments 
  of Dirac neutrinos derived in Ref.~\cite{Bel05}.
\end{thebibliography}
\end{document}